\documentclass[showpacs,pre,twocolumn,floatfix]{revtex4}

\usepackage{graphicx}
\usepackage{dcolumn}
\usepackage{bm}

\usepackage{amssymb}
\usepackage{amsmath}
\begin{document}

%\draft

\title{Inhomogeneous soliton ratchets under two ac forces}

\author{Luis Morales--Molina}
 \affiliation{Max-Planck Institut f\"ur Physik Komplexer Systeme,
N\"othnitzer Str. 38, 01187 Dresden, Germany.} 
%\email{Luis.Morales-Molina@uni-bayreuth.de}
\author{Franz G.\ Mertens}
\email{Franz.Mertens@uni-bayreuth.de} \affiliation{Physikalisches
Institut, Universit\"at Bayreuth, D-85440 Bayreuth, Germany}

\author{Angel S\'anchez}%
\homepage{http://gisc.uc3m.es/~anxo}
\affiliation{%
Grupo Interdisciplinar de Sistemas Complejos (GISC) and
Departamento de Matem\'aticas, Universidad Carlos III de Madrid,
Avenida de la Universidad 30, 28911
Legan\'es, Madrid, Spain, and\\
Instituto de Biocomputaci\'on y  F\'{\i}sica de
Sistemas Complejos (BIFI),\\
Facultad de Ciencias, Universidad de Zaragoza, 50009 Zaragoza, Spain}

\date{\today}


\begin{abstract}
We extend our previous work on soliton ratchet devices
[L.\ Morales-Molina {\em et al.}, Eur.\ Phys.\ J.\ B {\bf 37}, 79 (2004)]
to consider the joint effect of two ac forces
including non-harmonic drivings, as proposed for particle 
ratchets by Savele'v {\em et al.} 
[Europhys.\ Lett.\ {\bf 67}, 179 (2004);
Phys. \ Rev. \ E {\bf 70} 066109 (2004)].
Current reversals due to the interplay between the phases,
frequencies and amplitudes of the harmonics are obtained.
An analysis of the effect of the damping coefficient on the dynamics
is presented. We show that solitons give rise to non-trivial differences
in the phenomenology reported for particle systems that arise from their
extended character. A comparison with soliton ratchets in homogeneous systems
with biharmonic forces is also presented. 
This ratchet device may be an ideal candidate for
Josephson junction ratchets with intrinsic large damping.      
\end{abstract}

\maketitle

The understanding of ratchet mechanisms is a very active field,
of wide interest by its potential application in the design of
devices with new transport properties. The key feature of
ratchet devices is their ability to rectify the motion of
particles subjected to an external ac force with zero time average.
Originally proposed as a toy model for molecular motors, 
in the last decade many proposals have been put forward for
devices that use this ratchet phenomenon in different applications
\cite{reviews}. 
Ratchets working with extended particles (solitons) were 
subsequently studied as a generalization of particle ratchets
\cite{yuri}. Recently, a ratchet device driven by two ac 
forces was introduced by Savele'v and coworkers \cite{savelev},
in which the combination of the two drivings produced a variety 
of interesting phenomena and allowed a finely tunable control. 
In view of the rich behavior demonstrated by this system, a very
natural issue is its extension to soliton ratchets, that have 
important applications which can benefit from this proposal. 

Soliton ratchets under the presence of two ac forces have 
been extensively studied \cite{flach,Salerno-mixing,Niur} in homogeneous
systems, i.e., without an underlying ratchet potential. In this case,
it has been shown that the
ratchet mechanism works only for asymmetric
biharmonic forces, where the appearance of a directed translational
motion is a result of the effective coupling between the internal
mode (oscillations of the soliton width) and the external driving
force. In this paper, we focus on a soliton ratchet device recently 
studied by us \cite{pub,nuevo}, based on a nonlinear Klein-Gordon 
model with a lattice of delta-like inhomogeneities that induce a 
ratchet potential for the solitons.  
As we will see, one of the advantages of our model
as compared to the homogeneous system is 
that it works irrespectively of the symmetry of the ac driving: 
Directional transport takes place for ac forces with 
commensurate frequencies. Moreover, 
in the homogeneous system the motion drastically decreases for increasing damping,
due to the slowing down of the soliton width oscillations \cite{Niur}, while in the 
present system the ratchet phenomenon is present up to rather large 
values for the damping.

Our model, first introduced in \cite{pub}, is given by 
\begin{equation}\label{damp}
\ \phi_{tt}+\beta\phi_{t}-\phi_{xx}+ \sin(\phi)[1+V(x)]=f(t),
\end{equation}
where $f(t)=A_{1} \sin(\omega_{1} t+\theta_{1}) + A_{2} \sin( \omega_{2} t + \theta_{2})$, being $A_{1}$, $A_{2}$ the respective amplitudes of forces, $\omega_{1}$, $\omega_{2}$ the frequencies and $\theta_{1}$ and $\theta_{2}$ the phases of the harmonics (see e.g. \cite{savelev}).
Here we focus on the sine-Gordon model, although the same scheme can also
be applied to the general nonlinear Klein-Gordon model \cite{nuevo}.
For $V(x)$, we choose a spatially periodic potential, where the unit cell
is given by an asymmetric array of delta functions 
(inhomogeneities) in order to produce a ratchet-like phenomenon.
Specifically, the unit cell, of length $L$, is defined by three inhomogeneities
with the same intensity, the first one located at the beginning of
the cell, the second one at a distance $a$ from the first one, and
the third one at a distance $b$ from the second one, i.e.,
\begin{eqnarray}
V(x)=\epsilon \sum_{n}\left[\delta(x-x_{1}-nL)+\delta(x-x_{2}-nL)\right.\nonumber\\
\left.+\delta(x-x_{3}-nL)\right],
\end{eqnarray}
where  $L=a+b+c$, $a=x_{2}-x_{1}$, $b=x_{3}-x_{2}$
and $c=L+x_{1}-x_{3}$, with $x_{1}<x_{2}<x_{3}$. The parameters $(a,b,c)$ are chosen to be comparable to the
static soliton width $l_{0}$ in absence of inhomogeneities. In
addition, these should fulfill the conditions $a,b<c$
with $a\neq b$.

\begin{figure}[t]
\begin{center}
\includegraphics[width=8.5cm,angle=0]{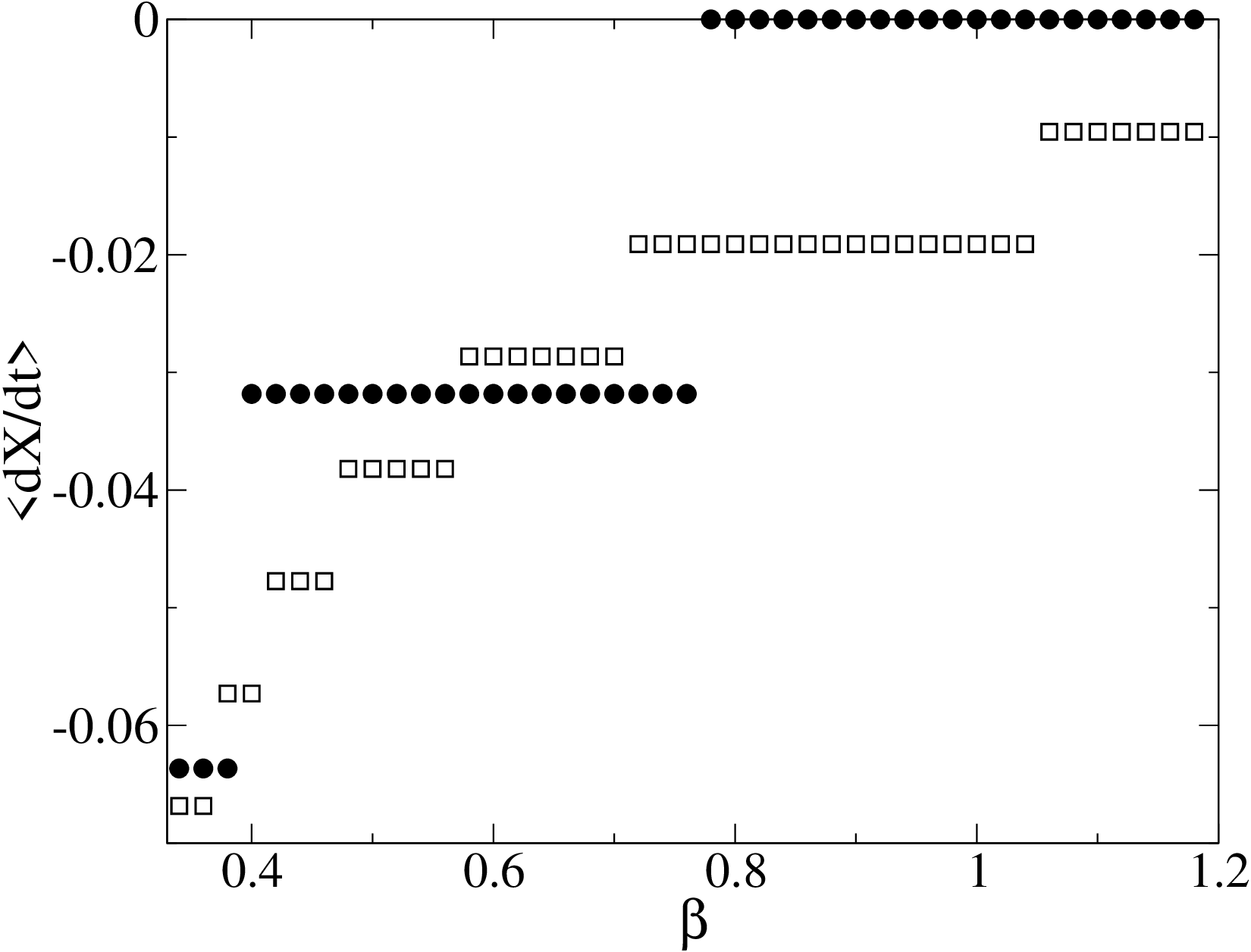}
\end{center}\vspace{-0.5cm}
\caption{Mean velocity vs damping coefficient $\beta$ for
different frequencies.  squares: $\omega=0.015$; filled circles:
$\omega=0.05$. The parameters used are $\epsilon=0.5$, $A=0.2$,
$x_{1}=0.5$, $x_{2}=1$, $x_{3}=2.3$ and $L=4$.} \label{dampfig}
\end{figure}

As a preliminary step,
we begin our study by analyzing how the system works in different 
 regimes of damping. This is very relevant to applications such as 
Josephson junctions. While the standard Josephson
junctions work usually at small damping, junctions with
intrinsically high damping such as superconductor insulator
normal-conductor insulator superconductor LJJ or high-$T_{c}$ LJJ
technology can also be fabricated \cite{koelle}.
For this particular aspect, we look at 
the case of a single harmonic component in Eq.\ (\ref{damp}).
In this case, according to \cite{pub,nuevo} we have a directed motion 
whose direction is determined by the position of the inhomogeneities,
and the dynamics reduces to a system very much like a rocking ratchet for
a single particle \cite{hangi}. Our results are shown in 
Fig.\ \ref{dampfig}, where  
we have chosen a lattice of inhomogeneities whose configuration
($a<b$) yields a negative current \cite{pub}. Hereafter $\langle \dot{X}
  \rangle $ means the time average of the velocity.
The figure demonstrates the 
increment of the efficiency upon decreasing the
damping $\beta$; however, nonzero values for the velocity are found for rather large
damping values, and for the lower frequency there is still motion even for $\beta>1$.

We note also that for these frequencies with lower damping values, the dynamics 
results depend on the initial conditions, which may lead to a chaotic
dynamics.

\begin{figure}[thp]
\begin{center}
\includegraphics[width=8.5cm,angle=0]{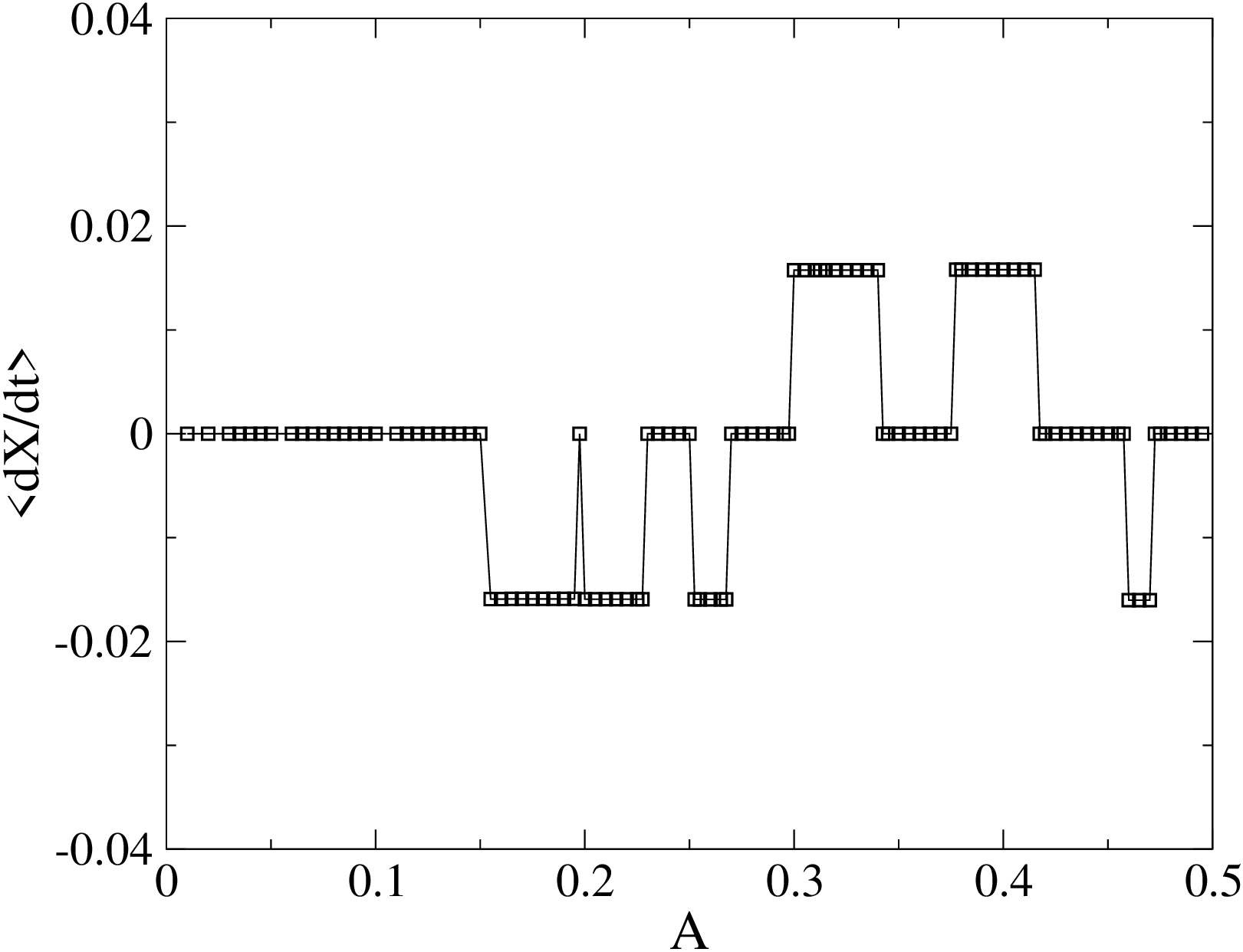}
\end{center}
\vspace{-0.5cm}
\caption{Mean velocity vs force amplitude $A=A_{1}=A_{2}$
Simulations of Eq.\ (\ref{damp}); the parameters used are $\beta=1$,
$\omega_{1}=0.025$, $\omega_{2}=0.3$, $\epsilon=0.5$, $\theta_{1}=\theta_{2}=0$,
$x_{1}=0.5$, $x_{2}=1$, $x_{3}=2.3$ and $L=4$. The solid line is a guide for the
eye.} \label{motion-beta=1}
\vspace{-0.5cm}
\end{figure}

Let us now move to the situation with two simultaneously acting ac forces.
For the time being, we work with sinusoidal forces as in Eq.\ (\ref{damp}), 
and we will consider later the case of rectangular pulses as in 
\cite{savelev}.
With two harmonics present (doubly rocked ratchet) we have a system in which
the symmetry can be broken both spatially (reflection symmetry $V(x)=V(-x)$) and
temporally (time shift symmetry $f(t)=-f(t+T/2)$ with $T=2\pi/\omega_{1}$). As we will now see, 
it is possible to obtain current reversals irrespective of the symmetry of the
biharmonic force.  The reason is that the underlying mechanism of transport
studied for homogeneous systems  subjected to biharmonic forces, 
that required temporal symmetry breaking \cite{Niur}, is not responsible for the directional
transport in our present model, as can be inferred from the fact that in the
inhomogeneous system we have directed transport under large damping. 
We examine the dynamics of Eq.\ (\ref{damp}) in the partially
adiabatic regime  $\omega_{1}/\omega_{2}\ll 1$ with $\omega_{1}\ll 1$ and
$\omega_{2}<1 $, setting $A_1=A_2$. In this case,
multiple current reversals are possible in the particle ratchet system \cite{savelev}.
Figure \ref{motion-beta=1} confirms clearly the existence of several
current reversals in our soliton device. The results show that we can reverse
the direction of motion by changing the force amplitudes, which, as in the 
particle case, opens the possibility to control the rectification properties 
in great detail. 

\begin{figure}
\begin{center}
\includegraphics[width=8cm,angle=0]{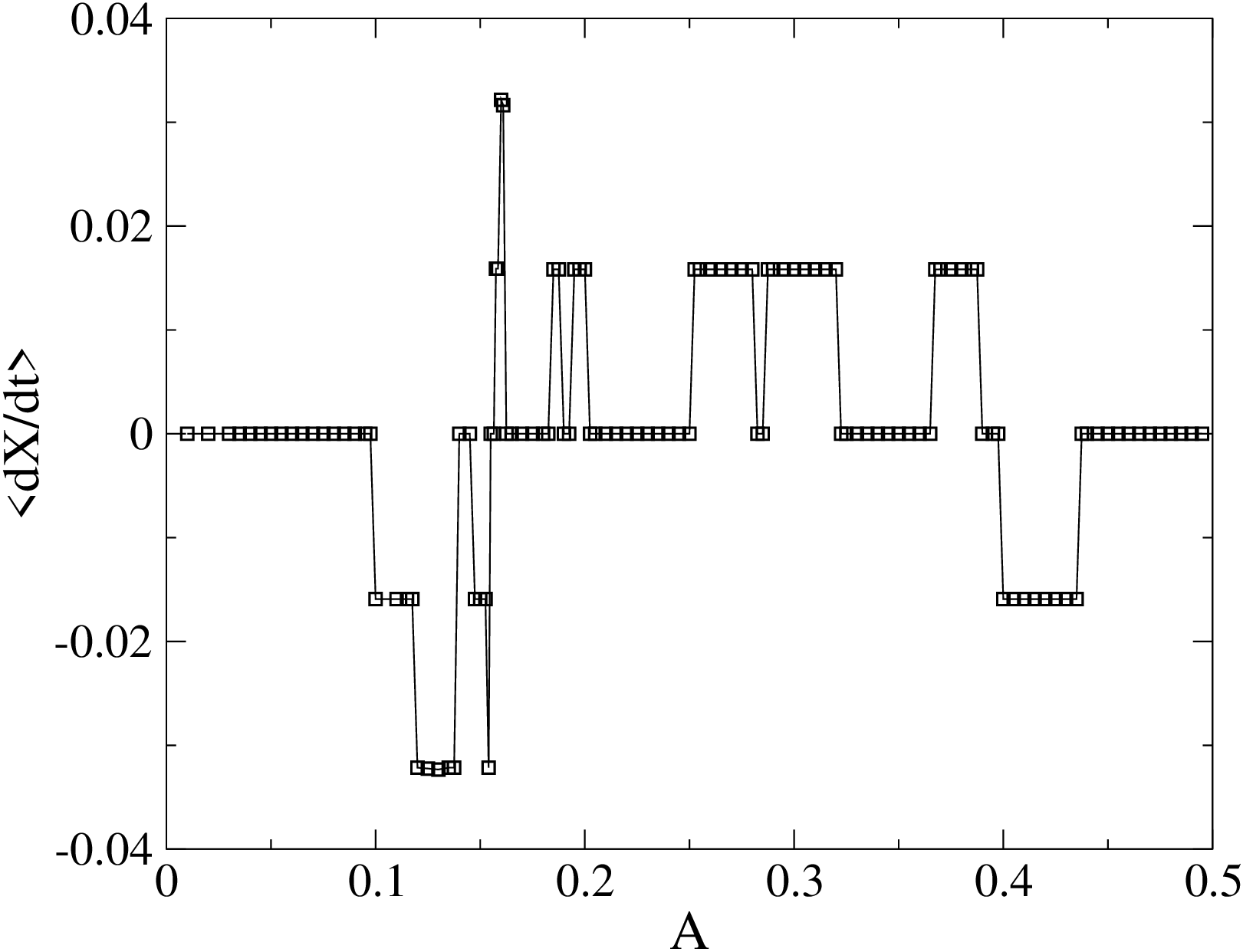}
\end{center}\vspace{-0.5cm}
\caption{Mean velocity vs force amplitude. $\beta=0.4$, $A_{1}=A_{2}=A$
simulations of Eq.\ (\ref{damp}). The other parameters are the same as in
Fig.\ref{motion-beta=1}. The solid line is a guide for the eye.} \label{motion-beta=0.4}
\end{figure}

In the previous analysis we have taken $\beta=1$, i.e., a rather large 
damping where the inertial effects are small. However, from the above
discussion for one harmonic, we expect changes in the behavior
also in the case of a biharmonic force when the damping is reduced.   
Figure \ref{motion-beta=0.4} exhibits indeed a 
different picture for the dynamics as compared with 
Fig.\ \ref{motion-beta=1}. The main differences are related to the shift
of the windows of motion towards lower force amplitude values as 
well as an increase of the absolute value of the mean
velocity for some force amplitudes. Interestingly, 
we note from this picture a non-typical behavior for the average
velocity close to the region where the first current reversal takes place:
We observe that, while for some values of the force amplitude the
windows of motion are suppressed, for other cases,
the absolute value of the average velocity is enhanced and even
reversed. 

\begin{figure}
\begin{center}
  \includegraphics[width=8.5cm,angle=0]{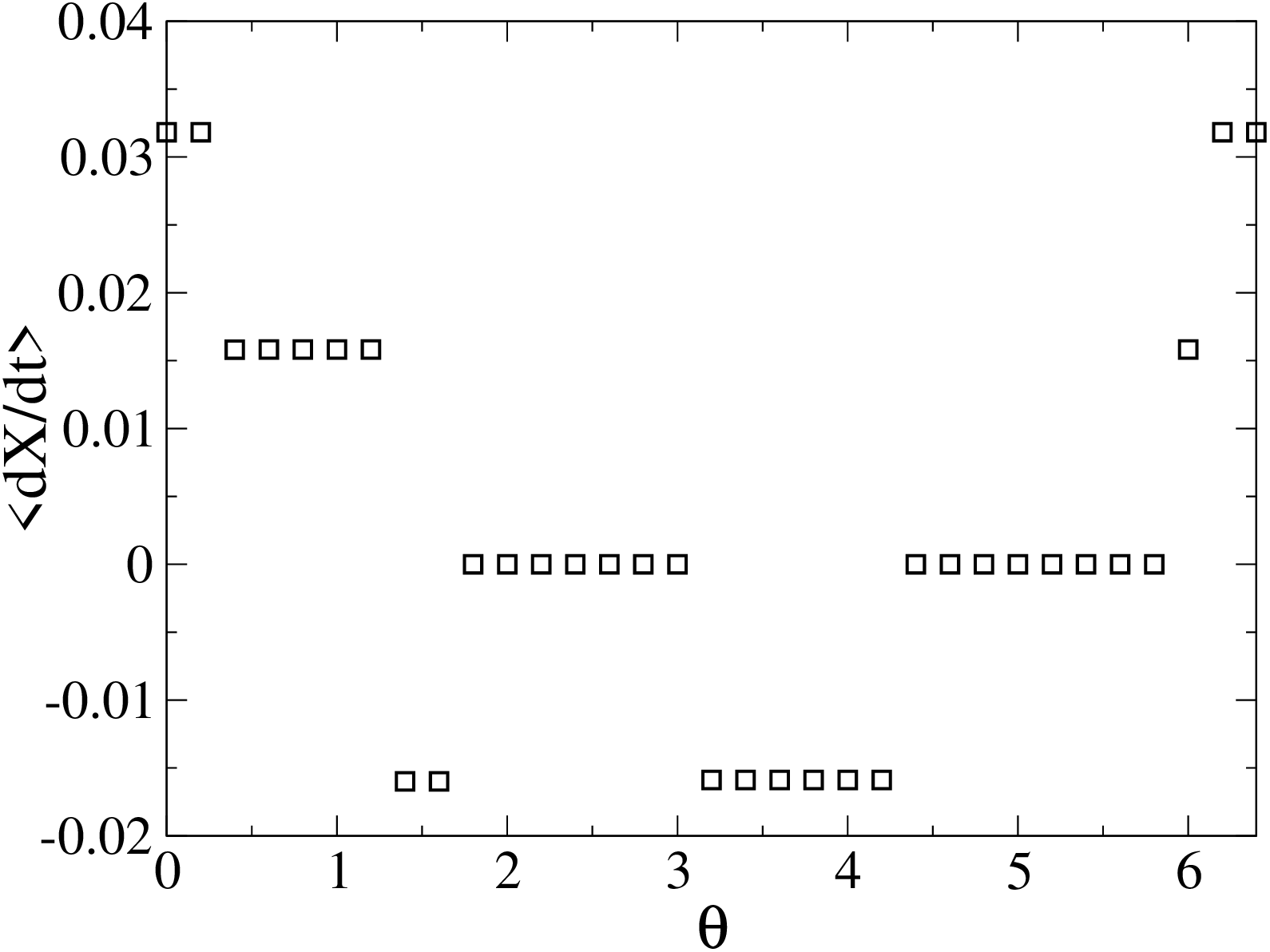}
\end{center}\vspace{-0.5cm}
\caption{Mean velocity vs $\theta$. Here we have taken $\theta_{1}=0$ and $\theta_{2}=\theta$ in Eq.\ (\ref{damp}). The other parameters are $A_{1}=A_{2}=0.16$, $\beta=0.4$, $\omega_{1}=0.025$ and $\omega_{2}=0.3$.}\label{phase-dependence}
\end{figure}

This issue can be further examined in Fig.\ \ref{phase-dependence},
where the dependence of the average velocity on the relative phase of the drivings 
is plotted. Here we see that by changing the phases we reverse the direction
of motion. This behavior is not only restricted to this singular region in Fig.\ref{motion-beta=0.4}; in fact,
it can be observed close to the regions where the currents are reversed in Fig.\ref{motion-beta=0.4}.
%By decreasing the damping coefficient this region, sensitive to changes in 
%the relative phase, approaches lower values of the force amplitudes.      
We stress that we have taken 
$\theta_1=0$ and $\theta_2=\theta$ without loss of generality, since although the dynamics obviously changes with both phases $\theta_1$ and $\theta_2$, one can always
map the choice for $\theta_1$ and $\theta_2$ into the previous representation
through the transformations
$\theta'_{2}=\theta_{2}-(\omega_{2}/ \omega_{1})\theta_{1}$ with $\theta'_{1}=0$.  
A feature of the behavior for this doubly rocked ratchet system, which differs
from the homogenous system case, is the quantization of the velocity dependence
on the phase (see Fig.\ \ref{phase-dependence} and cf.\ Fig.\ 2 in 
\cite{Niur}). This is yet another evidence that the
ratchet mechanisms at work in the homogeneous and inhomogeneous cases are not
the same. Nevertheless,  
despite this quantized nature, it is feasible to control the direction of 
motion by tuning the phases of the biharmonic force as in the homogeneous case.

\begin{figure}
\begin{center}
 \includegraphics[width=8.5cm,angle=0]{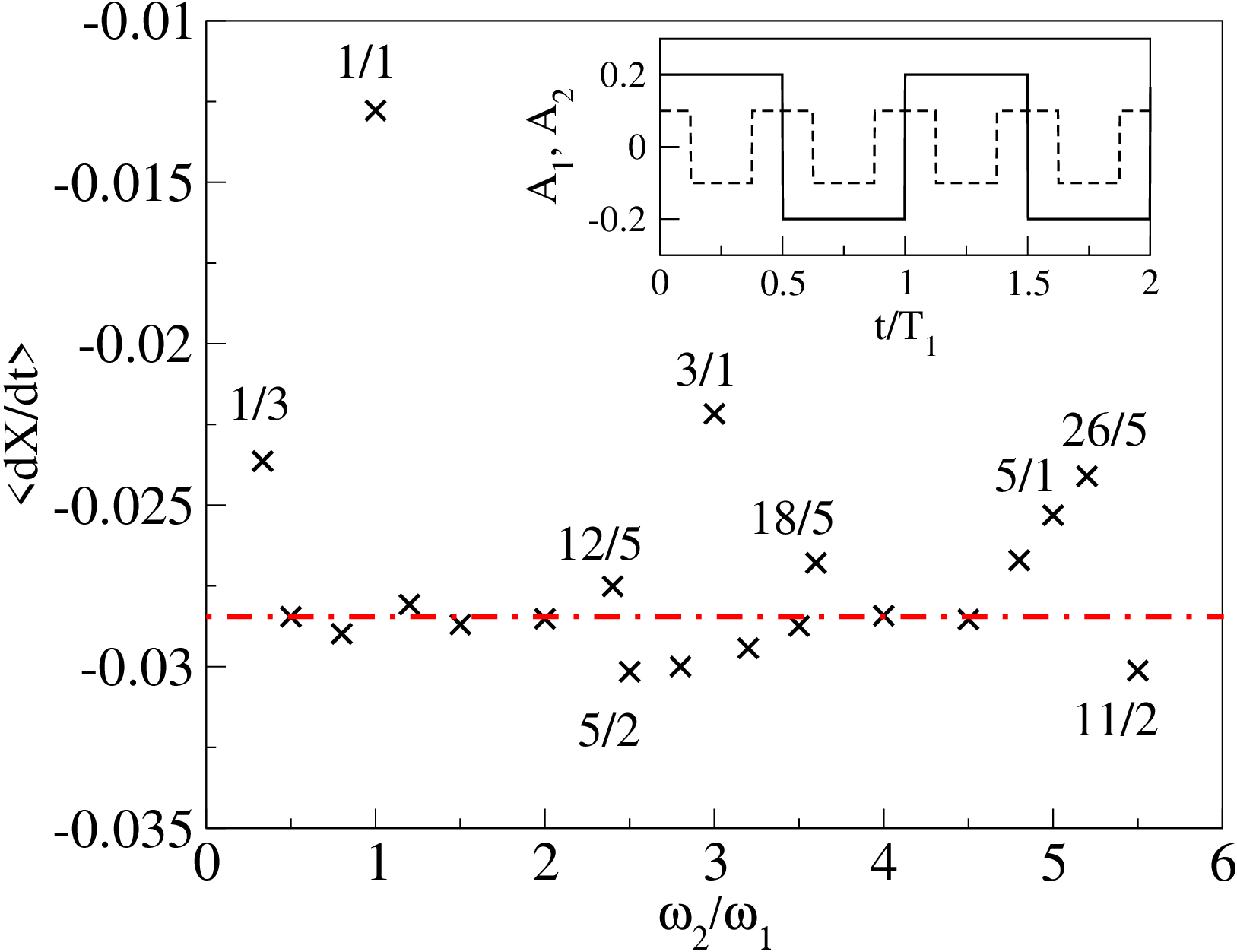}
\end{center}\vspace{-0.5cm}
\caption{(Color online) Mean velocity for different frequency ratios
  $\omega_{2}/\omega_{1}$ in the fully adiabatic regime 
  with rectangular wave signals
  $f(t)=A_{1}(t) +A_{2}(t)$. The
  parameters are $\omega_{1}=0.005$, $A_{1}=0.25$, $A_{2}=0.05$, $\beta=1$ and
  $\theta_{1}=\theta_{2}=0$. Inset: Rectangular wave signals:
  $A_{1}(t)=A_{1} \operatorname{sgn}[\sin(\omega_{1}t+\theta_{1})]$ (solid line);
  $A_{2}(t)=A_{2} \operatorname{sgn}[\sin(\omega_{2}t+\theta_{2})]$ (dashed line);
  $A_{2}=A_{1}/2=0.1$, $\omega_{2}=2\omega_{1}$, $\theta_{1}=0$ and $\theta_{2}=\pi/2$.  }\label{omega1/2-simulaciones}
\end{figure}

In view of the similarities with the 
single particle ratchet system in
\cite{savelev}, it is important to assess the degree of similarity between
the two systems. Savele'v and coworkers focus mainly on the case of 
rectangular
wave signals as the asymmetry and nonlinearity-induced mixing are  
separable  \cite{savelev}.
For such rectangular wave signals  in the fully adiabatic regime
  ($\omega_{1},\omega_{2}\ll 1$) with a sawtooth
ratchet potential, Savele'v {\em et al.} report changes in the dynamics only
for a relation between the frequencies given by 
$\omega_{2}/\omega_{1}=(2m-1)/(2n-1)$. As Fig.\ \ref{dampfig} shows, in our 
case the behavior turns out to be much more complicated.
The average velocity fluctuates around the  
value indicated by a horizontal line, but it does 
exhibit very many peaks. Most importantly, 
unlike the single particle picture shown in \cite{savelev}, here the peaks appear not only for
fractional harmonics $\omega_{2}/\omega_{1}=(2m-1)/(2n-1)$,  but also
for harmonics that fulfill the relations $\omega_{2}/\omega_{1}=(2m-1)/2n$ and 
$\omega_{2}/\omega_{1}=2m/(2n-1)$.

\begin{figure}[tp]
\begin{center}
 \includegraphics[width=8.5cm,angle=0]{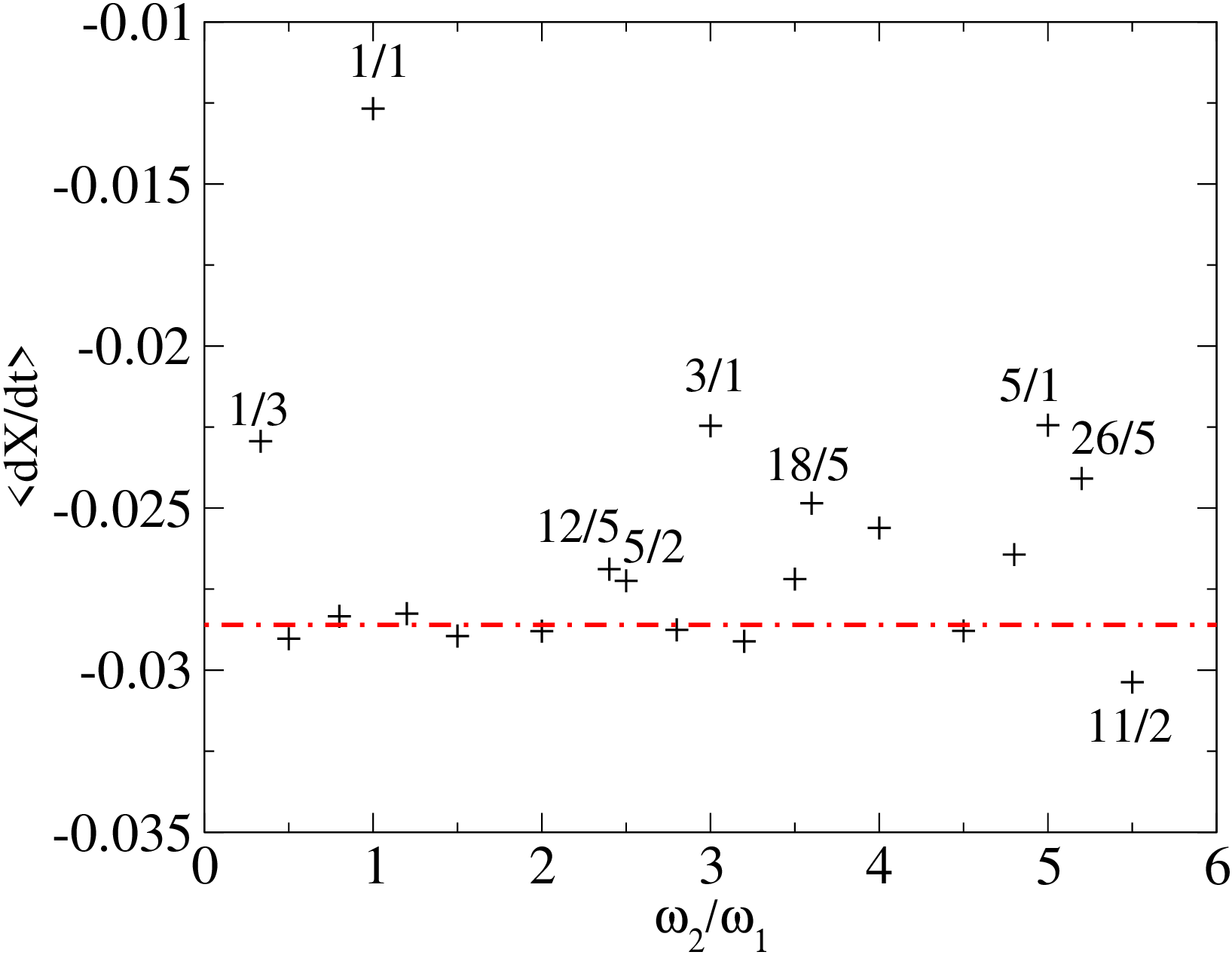}
\end{center}
\caption{(Color online) Collective Coordinates: Mean velocity for different frequency ratios $\omega_{2}/\omega_{1}$ in the fully adiabatic regime. $\omega_{1}=0.005$. The other parameters are $A_{1}=0.25$, $A_{2}=0.05$, $\beta=1$ and $\theta_{1}=\theta_{2}=0$.}\label{omega1/2-cc}
\end{figure}

The reason for the difference between the particle and the soliton doubly 
rocked ratchets can be traced back to the 
extended character of our solitons. 
It is well known that the soliton dynamics is affected by the
deformation of the soliton width that accompanies its motion along an array
of inhomogeneities \cite{nuevo}. This in turn modifies the 
effective potential arising in the description of the soliton as point 
particle, as well as in the variations of
the corresponding effective force. Accordingly, there is a large degree of
feedback between the soliton width and the soliton motion.
To obtain some insight on these issues, it is necessary to study the evolution
of the degrees of freedom that are involved. In order to do so, we resort 
to the use of the collective coordinate approach \cite{SIAM} which involves the 
soliton
width as an additional degree of freedom. In doing so, we find that 
the soliton can be 
described by two collective variables $X$ and $l$, whose
expressions are given by Eqs.\ (4)-(5) in \cite{nuevo}, with
the difference that now $f(t)$ contains the two rectangular wave signals 
whose expression appears in the 
caption of Fig.\ \ref{omega1/2-simulaciones}. This is in contrast to the
single particle doubly rocked ratchet of Savele'v {\em et al.}, because 
of the appearance of the width degree of freedom. 
This collective coordinate approach explains in full
detail the results of the simulations, as shown 
in Fig.\ \ref{omega1/2-cc} for the fully adiabatic regime. 
Again, as in the
simulations of Eq.\ (\ref{damp}) we note not only the existence of peaks for the
frequency ratios $\omega_{2}/\omega_{1}=(2m-1)/(2n-1)$, but also for some
frequencies ratio that fulfill the relation $\omega_{2}/\omega_{1}=(2m-1)/2n$
and $\omega_{2}/\omega_{1}=2m/(2n-1)$, absent in the single particle picture  \cite{savelev}.
This result is a clear demonstration of the role of the soliton width in 
the dynamics.

To conclude, we have generalized 
the results obtained for doubly rocked particle ratchets 
\cite{savelev} to extended systems, 
finding new phenomena that arise from the intrinsic width of the solitons. 
It was shown that the direction of motion can be modified by changing the
relation between the frequencies ratio, the phases of the harmonic forces
as well as their amplitudes.  However, in the frame of the fully adiabatic
regime of our soliton ratchet we find many
more peaks than in the particle ratchet system of Savele'v {\em et al.}, 
as we see special velocities for several types of frequency ratios. 
We have been able to show that this is thoroughly accounted for 
by the interplay of the soliton width and motion. 
We have also compared to doubly rocked soliton ratchets in homogeneous systems 
\cite{flach,Salerno-mixing,Niur} and verified that, although the soliton width
is involved in both cases, the mechanism for the appearance of the ratchet 
effect is different. Aside from the fact that homogeneous soliton ratchets 
arise only for asymmetric biharmonic drivings, further important differences
include the damping dependence of the velocity 
and the quantization of the dependence of the velocity on the relative 
phase. 
We emphasize that our ratchet system can be straightforwardly implemented in 
a Josephson junction device \cite{koelle}. In that case, the very many 
possibilities for the motion we have reported here would allow for a 
highly controllable device that can be tailored to fit different specific 
applications. The property that the ratchet phenomenon is observed even 
in the presence of large damping makes this system preferable to a 
homogeneous one driven by a biharmonic force, and makes it much more 
suitable for real life applications.
Finally, we note that an analysis similar to the present 
one can be 
extended to incommensurate ac forces with irrational values for the relation
$\omega_{2}/\omega_{1}$. However, our preliminary results show that this 
choice gives rise to new phenomena that require careful 
attention, and therefore it will be the 
object of a future investigation. 

 \*Acknowledgments: LMM thanks B. Lindner for comments.

\end{document}